\documentclass[preprint,12pt]{elsarticle}

\usepackage{amssymb}
\usepackage{amsmath}
\usepackage[version=4]{mhchem}
\usepackage{graphicx}
\usepackage{xcolor}
\usepackage{booktabs}
\usepackage{siunitx}
\usepackage{longtable}
\usepackage[utf8]{inputenc} 
\usepackage[T1]{fontenc}
 
\journal{Fusion Engineering and Design}

\begin{document}

\begin{frontmatter}



\title{Dynamic analysis and control design for the gas distribution
and storage system of the tritium fuel cycle in EU-DEMO}

\author[label1,label2]{C.A. Gómez-Pérez\corref{cor1}} 
\ead{c.a.gomez.perez@tue.nl}
\cortext[cor1]{Corresponding author}
\author[label1,label2]{M. van Berkel} 
\ead{M.vanBerkel@differ.nl}
\author[label1]{L. Özkan} 
\ead{l.ozkan@tue.nl}

\affiliation[label1]{organization={Eindhoven University of Technology, Department of Electrical Engineering, Control Systems},
         addressline={De Zaale},
         city={Eindhoven},
         postcode={5612 AE},
         country={Netherlands}}
\affiliation[label2]{organization={Dutch Institute For Fundamental Energy Research},
        addressline={De Zaale 20},
        city={Eindhoven},
        postcode={5612 AJ},
        country={Netherlands}}

\begin{abstract}

This work is concerned with the development of control strategies for the Direct Internal Recycling Loop (DIRL) system, which is an essential part of the Tritium Fuel Cycle (TFC) for the fueling of fusion reactors. As a first step, a control-oriented model is developed that describes dynamic behavior of DIRL with interactions between the torus reactor, buffer and fuel units, and recirculation streams. This model is used to evaluate the controllability, stability of the DIRL, and interactions between input and output variables. Moreover, the direct recycling of isotopes from the exhaust gases is discussed from a control perspective. It is observed that Gas Distribution and Storage (GDS) within the DIRL is associated with significant control challenges due to input–output interactions and competing process objectives. Three control strategies are developed and evaluated for the GDS of the European Demonstration fusion power plant (EU-DEMO): a Multiple Input Multiple Output (MIMO) control scheme, a redesigned GDS configuration with extended input variables enabling decentralised Single Input Single Output (SISO) control, and an extended-input MIMO control scheme addressing protium dilution. All strategies are assessed against three control objectives: maintaining GDS pressure around a prescribed set-point to ensure process safety; regulating the tritium–deuterium fuelling ratio for optimal reactor operation; and managing protium concentration to prevent fuelling dilution.

\end{abstract}

\begin{graphicalabstract}
\end{graphicalabstract}

\begin{highlights}
\item A control-oriented dynamic model for direct inner circulation loops is proposed, supporting both dynamic analysis and control design.
\item Direct recirculation of exhaust gases reduces tritium draw from storage, while steady-state protium composition at the fusion reactor inlet increases with the separation yield of the metal foil and vacuum pump unit.
\item Relative Gain Array (RGA) analysis is used to identify the preferred input-output pair structure selection for the direct inner circulation loop.
\item Based on RGA analysis, an actuator placement strategy is proposed that reduces input–output interactions and improves decentralised SISO control performance.
\item An LQR MIMO controller is also designed achieving good closed-loop performance; however, protium build-up cannot be avoided with a limited input set.
\item An adjusted LQR MIMO controller with an extended input set is proposed to address the protium dilution objective and overcome protium build-up
\end{highlights}

\begin{keyword}
Tritium Fuel Cycle \sep DEMO \sep Direct Inner Recirculation Loop \sep Hydrogen gas dynamics \sep Multivariable control \sep Optimal actuator selection
\end{keyword}

\end{frontmatter}



\section*{Nomenclature}

\begin{longtable}{@{}ll@{}}
\toprule
\textbf{Symbol} & \textbf{Description} \\
\midrule
\endhead

\multicolumn{2}{@{}l}{\textit{Roman symbols}} \\[2pt]
$N$       & Mol \\
$n$       & number of elements \\
$\dot{n}$ & Mol flow \\
$t$       & time \\
$V$       & Volume \\
$P$       & Pressure \\
$T$       & Temperature \\
$R$       & Ideal gas constant \\
$F$       & Volumetric flow \\
$x$       & Mol fraction \\
$C$       & Mol concentration \\
$\textbf{x}$ & State variable vector \\
$\textbf{u}$ & Input variable vector \\
$\textbf{y}$ & Output variable vector \\
$C_d$     & Valve coefficient \\
$A$       & Valve throughput area \\
$r$       & Reaction rate \\
$k$       & Kinetic coefficient \\
$K$       & Gas isotopic equilibrium constant \\
$a$       & Arrhenius parameter \\
$b$       & Arrhenius parameter \\
$IR$      & Isotopic ratio \\
$r_y$     & Output reference \\
$K_x$     & State feedback matrix \\
$K_i$     & Integral term matrix \\
$Q_x$     & Weighted matrix for state deviation term \\
$R_u$     & Weighted matrix for control effort term \\ [6pt]

\multicolumn{2}{@{}l}{\textit{Greek symbols}} \\[2pt]
$\rho$   & Molar Density \\
$\eta$   & Separation yield \\
$\kappa$ & Transfer function gain \\
$\tau$   & Transfer function time constant \\
$\alpha$ & Valve aperture \\
$\omega$ & Mass weight \\
$\gamma$ & Specific heat ratio \\
$\lambda$ & Eigenvalue \\
$\Lambda$ & RGA matrix \\ [6pt]

\multicolumn{2}{@{}l}{\textit{Subscripts}} \\[2pt]
$_g$     & Gas \\
$_{atm}$ & Atmospheric \\
$_{FV}$  & Feed Vessel \\
$_{BV}$  & Buffer Vessel \\
$_{FV1}$ & Fuel Vessel 1 \\
$_{FV2}$ & Fuel Vessel 2 \\
$_{PI}$  & Pellet Injection \\
$_{GP}$  & Gas Puffing \\
$_{GI}$  & Gas Injection \\
$_{Ar}$  & Argon \\
$_{Xe}$  & Xenon \\
$_{He}$  & Helium \\
$_i$     & component $i=[H_2,D_2,T_2,HD,HT,DT, \ce{^3He},NG]$ \\
$_j$     & isotopes $j=[T,D,H, \ce{^3He},NG]$ \\
$_H$     & Protium \\
$_D$     & Deuterium \\
$_T$     & Tritium \\
$_{H_2}$ & Hydrogen \\
$_{NG}$  & Noble Gas \\
$_{crit}$ & Critical \\
$_{s}$   & Subsonic flow \\
$_{c}$   & Choked flow \\
$_{tok}$ & Tokamak reactor \\
$_{exh}$ & Gas exhaust \\
$_F$     & Frobenius \\[6pt]

\multicolumn{2}{@{}l}{\textit{Acronyms}} \\[2pt]
TFC      & Tritium Fuel Cycle \\
DEMO     & Demostration Power Plant \\
EU-DEMO  & European Demonstration Power Plant \\
DIRL     & Direct Internal Recycling Loop \\
INTL     & Inner Tritium Plant Loop \\
OUTL     & Outer Tritium Plant Loop \\
GDS      & Gas Distribution and Storage \\
SOL      & Scrape-Off Layer \\
RGA      & Relative Gain Array \\
DT       & Deuterium and Tritium \\ 
ODE      & Ordinary Differential Equations \\
SISO     & Single Input Single Output \\
MIMO     & Multiple Input Multiple Output \\
LQR      & Linear Quadratic Regulator \\
PID      & Proportional-Integrative-Derivative control \\ [6pt]

\bottomrule
\end{longtable}

\section{Introduction}
\label{sec1}

The international research roadmap for fusion energy development focuses on the development of DEMO (Demonstration Power Plant) reactors, intended to demonstrate the commercial viability of fusion energy. DEMO facilities will move beyond scientific demonstration by generating net electricity, operating with high reliability, and testing technologies required for industrial deployment. One of the key research priorities is to achieve a sustainable use of tritium as fuel. Hydrogen gas composed of tritium and deuterium (DT) isotopes fuels the fusion reactions in future fusion reactors, stellarators, and tokamaks alike. The chemical process plant known as Tritium Fuel Cycle (TFC) has been proposed for the European Demonstration Power Plant (EU-DEMO) as a process that can extract unburnt tritium from exhaust gases and secondary enclosures, integrate tritium from the breeding blankets, and maintain the isotopic composition \cite{Day2022}.

Tritium self-sufficiency is essential to making fusion energy feasible. Because tritium is not readily available as a resource, it must be artificially produced in the fusion reactor using breeding blanket modules \cite{Moral2013}. Self-sufficiency will be possible with appropriate tritium breeding and management, minimizing residence time throughout the entire TFC chemical plant \cite{Abdou2021}. In Day et al. \cite{Day2019}, the three-loop recirculation chemical plant is proposed. Such a concept refers to closed recirculation processes that extract exhausted tritium from the fusion reactor. Loop recirculation process are known as (Figure \ref{fig1}): 1. Direct Internal Recycling Loop (DIRL), 2. Inner Tritium Plant Loop (INTL), and 3. Outer Tritium Plant Loop (OUTL). DIRL uses a fast separation of hydrogen gas from the tokamak exhaust. INTL removes protium isotopes to deliver a concentrated DT mixture gas to the fusion reactor. and OUTL develops an exhaustive tritium recovery system to reduce pollutant emissions and incorporate tritium from breeding blankets to obtain a gas with a high tritium isotope concentration \cite{Day2019}. This work will focus on the DIRL chemical plant. The main goal of DIRL design is to deliver all the isotopes more quickly by directly separating hydrogen gas from the fusion reactor exhaust using metal foils and vacuum pumps \cite{Abdou2021} \cite{Day2019}. Thereby minimizing the tritium inventory. This is crucial for the economic viability of the tritium plant, given the cost of producing and breeding tritium. DIRL comprises the fusion reactor, the metal foil, the vacuum pump, the fueling system (pellet injection and gas puffing), and the GDS system (see Figure \ref{fig2})  
 
The main purpose of the GDS, part of the DIRL, is to manage recovered tritium, either storing it or using it as fuel in the Tokamak fusion reactor. The GDS system comprises 4 compressed gas vessels \cite{Day2022}. These vessels should remain in safe operating condition, \textit{i.e.}, maintain safe pressure operation. The isotopic composition of the fuel delivered to the fusion reactor should be maintained at 50/50 deuterium-tritium (DT) to optimize energy generation. Moreover, inert components like protium should be diluted to minimize fuel dilution and radiation losses; fuel protium content should be below $1\%$ \cite{Day2022}. The previously mentioned operational requirements should be achieved through the control systems. An appropriately designed control scheme is necessary to ensure a safe and reliable fuel cycle system that maintains conditions for tritium self-sufficiency. This is a well-designed control synthesis with appropriate dynamic analysis to guarantee stability and good performance.

The literature shows some advances in TFC design and technology selection (e.g., \cite{Day2019}; \cite{Lord2024}). However, no systematic dynamic analysis and control evaluation was performed. In this work, we first develop a control-oriented dynamic model of the DIRL. The model is then used to study interactions between units and gas streams, which is essential to develop control schemes. The model is also used to evaluate the GDS control system's capability to manage the protium composition of the gas fuel, which should be less than $1\%$. This is particularly difficult because direct hydrogen recirculation leads to protium build-up. 

Three control schemes are developed and assessed. 1. LQR control; 2. SISO control with input selection; 3. LQR control with extended input variables. We evaluate controllability, stability with dedicated tools and multivariable dynamic interactions with the Relative Gain Array (RGA) \cite{koenders2023}. Each scheme is evaluated against the achievement of three control objectives: 1. Safety by pressure control, 2. Maintain tritium-deuterium isotopic ratio for optimal reaction, and 3. Maintain low protium composition. 

The paper is structured as follows. Section \ref{sec2} develops the control-oriented dynamic model of the DIR loop. Section \ref{sec3} presents the controllability and interaction analysis, followed by the design of the proposed control schemes. Simulation results are reported in Section \ref{sec4}, and conclusions are drawn in Section \ref{sec5}.

\section{Control-Oriented Dynamic Modeling of DIRL}
\label{sec2}

\begin{figure}[t]
\centering
\includegraphics[scale=0.3]{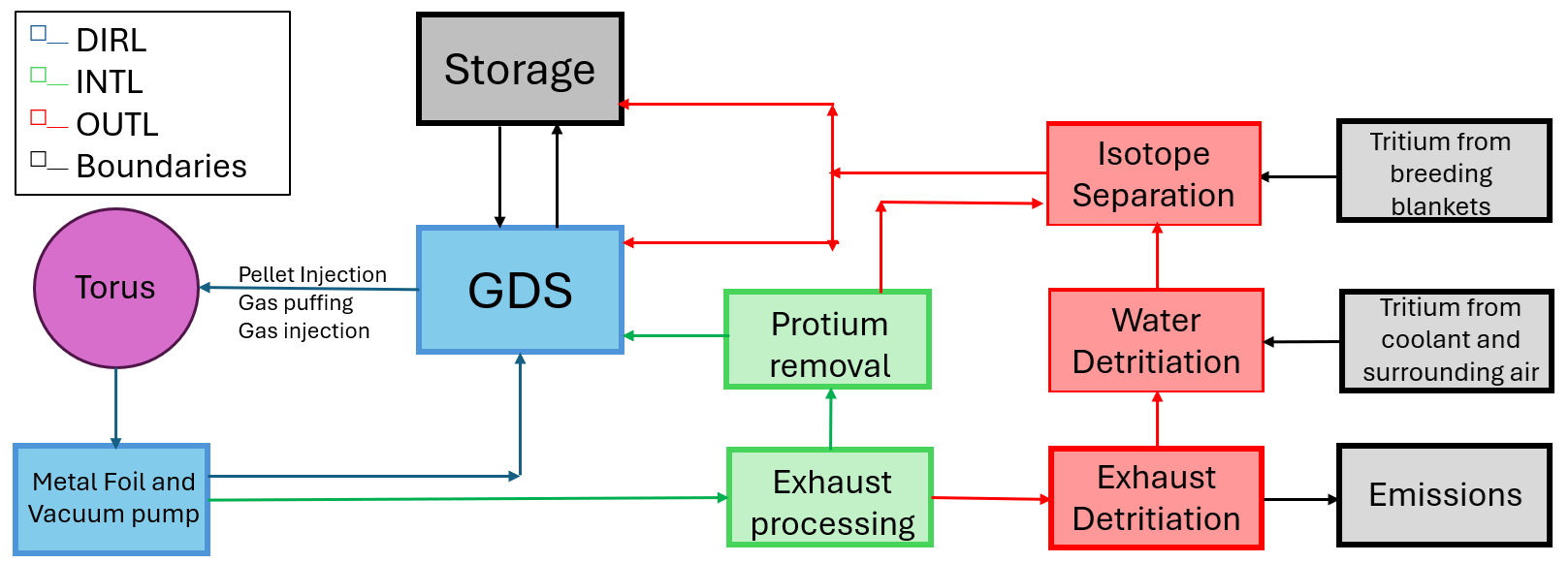}
\caption{Tritium Fuel Cycle.}\label{fig1}
\end{figure}

The DIRL is a recirculating fuel concept designed to minimise tritium inventory throughout the entire TFC by recirculating hydrogen gas comprising protium, deuterium, and tritium isotopes \cite{Day2019}. As illustrated in Figure \ref{fig1}, the TFC is primarily concerned with the exhaustive separation of tritium from the exhaust gas stream. A key enabling technology is the rapid separation of hydrogen isotopes via a metal foil and vacuum pump unit, which has been shown to significantly reduce tritium inventory at the plant level \cite{Day2022}.


\begin{figure}[t]
\centering
\includegraphics[scale=0.4]{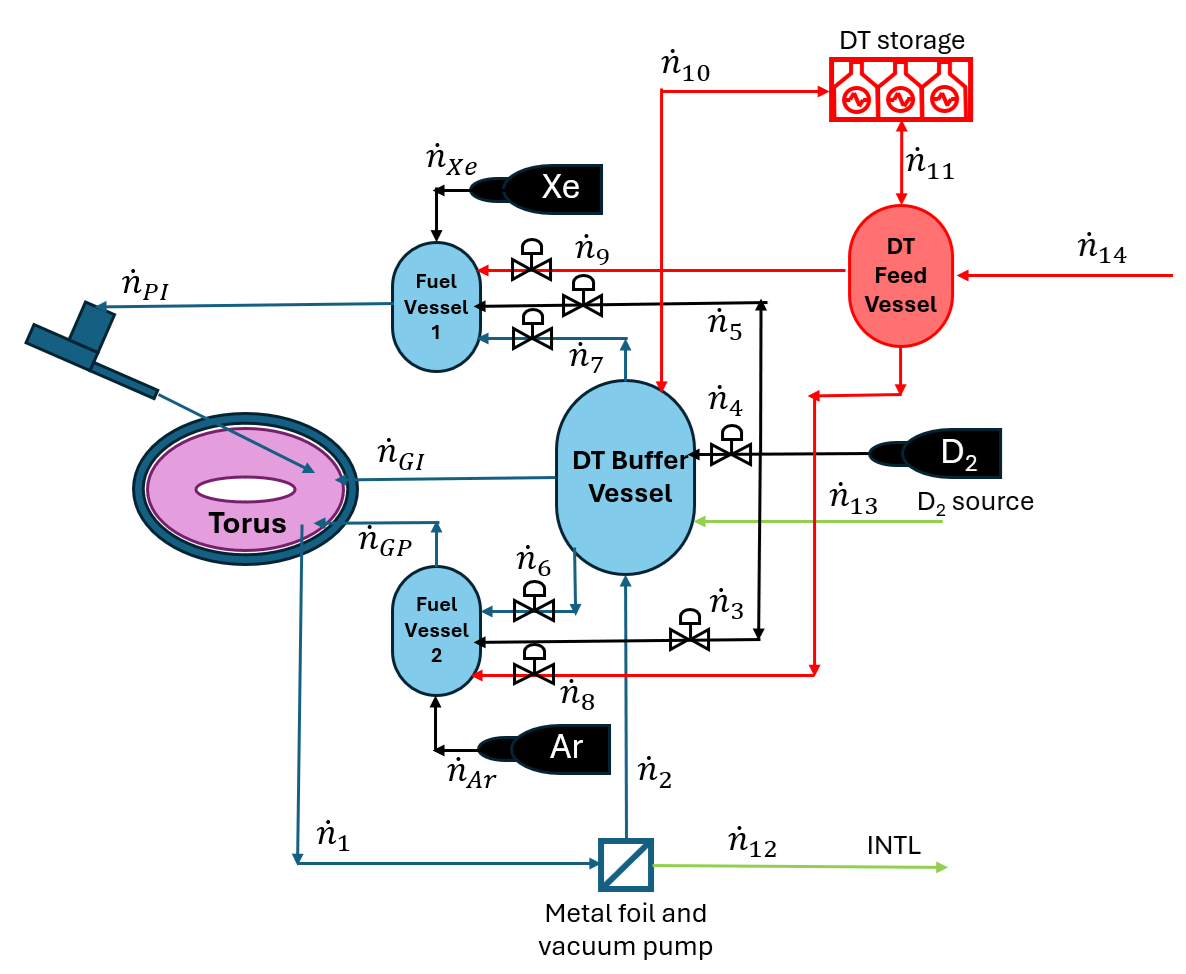}
\caption{DIRL Process Diagram.}\label{fig2}
\end{figure}

GDS is responsible for the hydrogen gas management. It comprises of 4 vessels (Figure \ref{fig2}): 1. fuel vessel 1 that feeds the pellet injection for fusion core fuelling, 2. fuel vessel 2 that feeds the gas puffing for edge plasma fueling, 3. Buffer vessel to receive hydrogen gas from the metal foil, INTL, and 4. DT storage vessel that receives highly tritium-concentrated gas from the OUTL.

The main goal of the model developed here is to assess the dynamic relationships among the possible isotopes of hydrogen gas (tritium, deuterium, and protium) and the dynamic characteristics of hydrogen management in the DIRL. Moreover, the model should be computationally light, enabling fast dynamic predictions for evaluating control designs. It is developed based on first principles (mass/mole and component balances) supported with data-driven parameter identification. The model captures dynamic interactions among four pressurized vessels, the torus reactor, the metal foil, the vacuum pump, and the DT storage system. Constant values, parameters, and operating conditions were directly obtained or identified from \cite{Jonas2024}.


\subsection{Gas Distribution and Storage System}
\label{subsec01}

\subsubsection{Pressure dynamics}
\label{subsubsec1}

Pressure dynamics is crucial for monitoring process safety operations. Therefore, total mol balances applied to each vessel (Figure \ref{fig2}) are used to develop the pressure dynamics. The total mol balance for the DT feed vessel is:

\begin{equation}
\frac{dN_{FV}}{dt}=\dot{n}_{14}+\dot{n}_{11}-\dot{n}_8-\dot{n}_9,
\end{equation}

Assuming that the gases exhibit ideal gas behavior, we can use the ideal gas law to obtain a relationship between moles and pressure ($PV=NRT$), where $P$ is the pressure, $V$ is the volume, $T$ is the operating temperature, which is assumed to be constant, and $R$ is the ideal gas constant. Molar flow is calculated using the molar gas density ($\dot{n}=\rho_gF$). Where $F$ is the volumetric flow, and $\rho_g$ is the molar gas density. We assume $\rho_g$ is constant around the operating conditions. We use the ideal gas law to predict the molar gas density using atmospheric pressure ($P_{atm}$) and operating temperature.

\begin{equation}
\rho_g=\frac{P_{atm}}{RT},
\end{equation}

Using previous relationships in the total mol balance equation, we can derive the differential equation for DT feed vessel pressure dynamics prediction:

\begin{equation}
\frac{dP_{FV}}{dt}=\frac{RT}{V_{FV}}(\dot{n}_{14}+\dot{n}_{11})-\frac{RT\rho_g}{V_{FV}}F_8-\frac{RT\rho_g}{V_{FV}}F_9,
\end{equation}

Therefore, by analogy, the differential equations used to predict the pressure dynamics of the other vessels can be derived in a similar manner:

\begin{equation}
\frac{dP_{BV}}{dt}=\frac{RT}{V_{BV}}(\dot{n}_{2}+\dot{n}_{4}+\dot{n}_{10}+\dot{n}_{13}-\dot{n}_{GI})-\frac{RT\rho_g}{V_{BV}}F_6-\frac{RT\rho_g}{V_{BV}}F_7,
\end{equation}

\begin{equation}
\frac{dP_{FV1}}{dt}=\frac{RT}{V_{FV1}}(\dot{n}_{5}+\dot{n}_{Xe}-\dot{n}_{PI})+\frac{RT\rho_g}{V_{FV1}}F_7+\frac{RT\rho_g}{V_{FV1}}F_9,
\end{equation}

\begin{equation}
\frac{dP_{FV2}}{dt}=\frac{RT}{V_{FV2}}(\dot{n}_{3}+\dot{n}_{Ar}-\dot{n}_{GP})+\frac{RT\rho_g}{V_{FV2}}F_6+\frac{RT\rho_g}{V_{FV2}}F_8.
\end{equation}

\subsubsection{Composition dynamics}
\label{subsubsec2}

Another important aspect of the dynamic is the inventory of important isotopes and gases that participate in Tokamak particle reactions, whether as reactants or products. This is necessary to predict the gas composition of all possible isotopic pairings. For the components' mol balances on each vessel (Figure \ref{fig2}), consider the subscript $i=[H_2,D_2,T_2,HD,HT,DT, \ce{^3He},NG]$, (Where $NG=\ce{^4He}+Ar+Xe$). Consider the components' mol balances on DT feed:

\begin{equation}
\frac{dN_{FVi}}{dt}=\dot{n}_{14i}+\dot{n}_{11i}-\dot{n}_{8i}-\dot{n}_{9i},
\end{equation}
considering that component moles can be calculated using molar fraction ($N_{i}=x_{i}N$ and $ \dot{n}_i=x_{i} \dot{n}$):

\begin{equation}
\frac{d(x_{FVi}N_{FV})}{dt}=x_{14i}\dot{n}_{14}+x_{11i}\dot{n}_{11}-x_{FVi}\dot{n}_{8}-x_{FVi}\dot{n}_{9},
\end{equation}

Since $N_{FV}$ is time-dependent, it is necessary to use the product rule for derivation

\begin{equation}
\frac{d(x_{FVi}N_{FV})}{dt}=x_{FVi}\frac{dN_{FV}}{dt}+N_{FV}\frac{dx_{FVi}}{dt},
\end{equation}

Moreover, using the ideal gas law equation and molar gas density definition (equation 2), it is possible to derive the differential equation that predicts the molar fraction dynamics in the DT feed vessel:

\begin{equation}
\begin{split}
\frac{dx_{FVi}}{dt}=\frac{RT}{V_{FV}P_{FV}}(x_{14i}\dot{n}_{14}+x_{11i}\dot{n}_{11})-\frac{RT\rho_g}{V_{FV}P_{FV}}x_{FVi}F_8 \\
-\frac{RT\rho_g}{V_{FV}}x_{FVi}F_9-\frac{x_{FVi}}{P_{FV}}\frac{dP_{FV}}{dt},
\end{split}
\end{equation}

Therefore, by analogy, the differential equations used to predict the fraction dynamics of the other vessels can be derived in a similar manner:

\begin{equation}
\begin{split}
\frac{dx_{BVi}}{dt}=\frac{RT}{V_{BV}P_{BV}}(x_{2i}\dot{n}_{2}+x_{4i}\dot{n}_{4}+x_{10i}\dot{n}_{10}+x_{13i}\dot{n}_{13}-x_{BVi}\dot{n}_{GI}) \\
-\frac{RT\rho_g}{V_{BV}P_{BV}}x_{BVi}F_6 -\frac{RT\rho_g}{V_{BV}P_{BV}}x_{BVi}F_7-\frac{x_{BVi}}{P_{BV}}\frac{dP_{BV}}{dt},
\end{split}
\end{equation}

\begin{equation}
\begin{split}
\frac{dx_{FV1i}}{dt}=\frac{RT}{V_{FV1}P_{FV1}}(x_{5i}\dot{n}_{5}+x_{NGi}\dot{n}_{Xe}-x_{FV1i}\dot{n}_{PI}) \\
+\frac{RT\rho_g}{V_{FV1}P_{FV1}}x_{BVi}F_7+\frac{RT\rho_g}{V_{FV1}P_{FV1}}x_{FVi}F_9-\frac{x_{FV1i}}{P_{FV1}}\frac{dP_{FV1}}{dt},
\end{split}
\end{equation}

\begin{equation}
\begin{split}
\frac{dx_{FV2}}{dt}=\frac{RT}{V_{FV2}P_{FV2}}(x_{3i}\dot{n}_{3}+x_{NGi}\dot{n}_{Ar}-x_{FV2i}\dot{n}_{GP}) \\
-\frac{RT\rho_g}{V_{FV2}P_{FV2}}x_{BVi}F_6-\frac{RT\rho_g}{V_{FV2}P_{FV2}}x_{FVi}F_8-\frac{x_{FV2i}}{P_{FV2}}\frac{dP_{FV2}}{dt},
\end{split}
\end{equation}

\noindent where $x_{NGi}$ only considers the composition of the noble gases, and it is zero for the other components. Notice that this is an implicit set of ODEs ($F(t,\textbf{x},\dot{\textbf{x}})$ where \textbf{x} is the state variables vector), since the molar fraction dynamics depend on the pressure derivative. These kinds of ODEs present particular numerical challenges. For instance, when the Jacobian $\partial F / \partial \dot{\mathbf{x}}$ is singular, the equation cannot be locally resolved for $\dot{\mathbf{x}}$ uniquely, admitting multiple solution trajectories from a given initial point. 
In this case, pressure dynamics do not depend on the molar fraction dynamics; therefore, numerical simulation is possible.

\subsubsection{Actuators modeling}
\label{subsubsec3}

The GDS system uses four types of actuators. 1. deuterium cylinder sources, 2. hydride beds, 3. valves between vessels, and 4. Metal foil and vacuum pump device separation yield, which results in hydrogen gas recirculation ($\dot{n}_2$). Cylinder sources for $D_2$ can easily manipulate the volumetric flow using a control valve; therefore, we will not consider a mathematical model for the input variables $\dot{n}_{3}, \dot{n}_{4},$ and $\dot{n}_{5}$. Hydride beds' storage will rely on the natural dynamics of chemisorption, which follow first-order kinetics \cite{sreeraj2022}. Therefore, hydride beds are modeled using a first-order transfer function \cite{seborg2016process} \cite{ljung1998system}:

\begin{equation}
\dot{n}(t)=\mathcal{L}^{-1}(N(s)),
\end{equation}

\noindent with

\begin{equation}
N(s)=\frac{\kappa U(s)}{\tau s+1}.
\end{equation}

The flow between pressurized vessels affects gas flow; to model this phenomenon, we use a valve equation. Since the TFC deals with hydrogen and noble gases, we will use the compressible isentropic valve equation to predict the volumetric flow. Hence, it is necessary to evaluate whether the flow is choked or not using the critical pressure ratio condition:


\begin{equation}
r_{crit}=\left(\frac{P_{down}}{P_{up}}\right)_{crit} = \left(\frac{2}{\gamma+1}\right)^{\frac{\gamma}{\gamma-1}},
\end{equation}

\noindent where $P_{down}$ is the lowest pressure, $P_{up}$ is the highest pressure and $\gamma$ is the specific heat ratio. The hydrogen value for $\gamma$ is 1.4 \cite{hilsenrath1956}. If the pressure ratio is higher than the critical pressure ratio, the flow is subsonic:

\begin{equation}
F_s=\frac{C_d \alpha A P_{up}}{\sqrt{\omega}\rho_g}\sqrt{\frac{2 \gamma}{RT(\gamma-1)}\left(\frac{P_{down}}{P_{up}}^{\frac{2}{\gamma}}-\frac{P_{down}}{P_{up}}^{\frac{\gamma+1}{\gamma}}\right)},
\end{equation}

\noindent if the pressure ratio is lower than the critical pressure ratio, we have choke flow:

\begin{equation}
F_c=\frac{C_d \alpha A P_{up}}{\sqrt{\omega}\rho_g}\sqrt{\frac{\gamma}{RT}}\left(\frac{2}{\gamma+1}\right)^{\frac{\gamma+1}{\gamma-1}}.
\end{equation}

Consequently, the calculated flow can transition between subsonic and choked conditions throughout the simulation. To avoid numerical problems, a smooth function is used to predict the volumetric flow:

\begin{equation}
F=\frac{1}{(\frac{1}{F_s^4}+\frac{1}{F_c^4})^{\frac{1}{4}}}.
\end{equation}

\subsection{Pellet Injection system}
\label{subsec02}

The pellet injector has no dynamic properties with respect to the tritium cycle. Hence, it will be modeled as a delay system. It takes 30 seconds from the gas inlet to the pellet injection for torus fuelling \cite{Jonas2024} \cite{Bosman2023}.

\subsection{Metal Foil and Vacuum Pump Device}
\label{subsec03}

It is assumed that all non-hydrogen gases go to the INTL. We also assume that the metal foil and vacuum pump device separation process is fast and always in a steady state. Thus, the system is evaluated using steady-state balances:

\noindent Total mol balance:
\begin{equation}
\dot{n}_1=\dot{n}_2+\dot{n}_{12}.
\end{equation}
Hydrogen mol balance:
\begin{equation}
x_{1H_2}\dot{n}_1=\dot{n}_2+x_{12H_2}\dot{n}_{12}.
\end{equation}
Noble gas mol balance:
\begin{equation}
x_{1NG}\dot{n}_1=x_{12NG}\dot{n}_{12}.
\end{equation}

Moreover, the separation yield ($\eta$) is a manipulated variable, ranging from 0 to the maximum achievable (80\%) \cite{Coleman2019}, and then $\dot{n_2}=\eta x_{1H_2}\dot{n}_1$.

\subsection{Fusion reactor}
\label{subsec04}

The torus is an important component of the process. In the context of the TFC, the Tokamak must be operated as a pulsed machine with four operational phases \cite{Jonas2024} \cite{Xu2010} \cite{Fitzpatrick2026}: 1). Ramp-up phase, where the gas is injected, and the plasma current starts, 2). Flat-top phase where the plasma reactions are performed, 3). Ramp-down phase, where the gas is extracted, and the central solenoid magnet starts to recharge, and 4) Dwell phase without plasma operation. This model is developed to evaluate the steady-state during the flat-top phase.

\begin{figure}[t]
\centering
\includegraphics[scale=0.4]{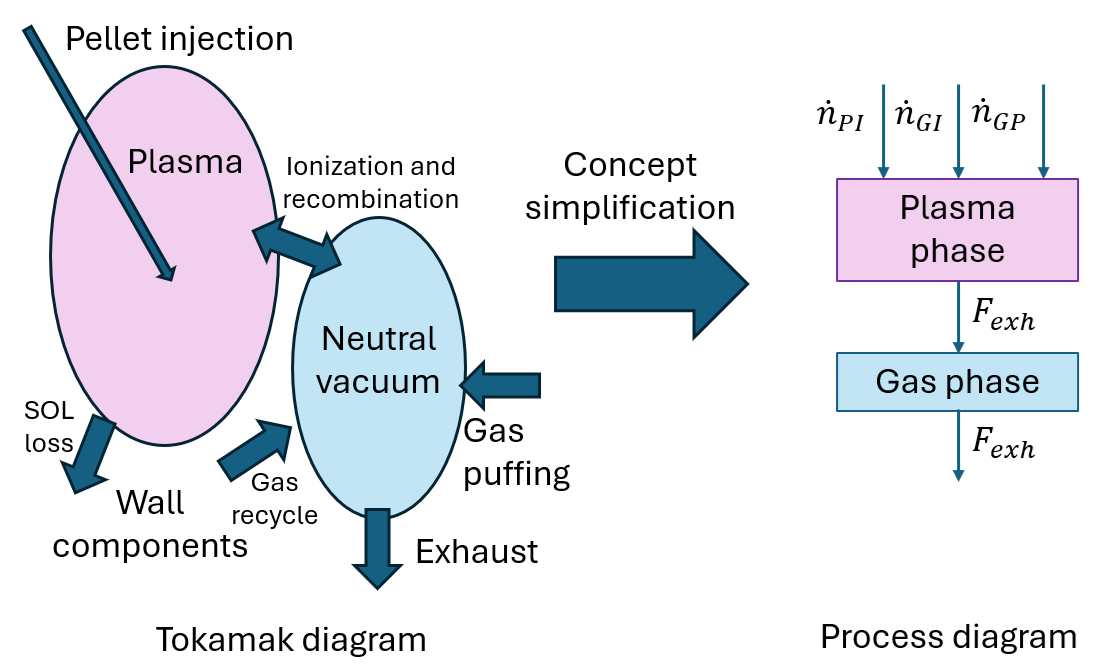}
\caption{Schematic representation of a tokamak fusion reactor and its process diagram concept simplification. Modified from \cite{Blanken2018}}\label{fig3}
\end{figure}

Generally, the torus is modeled using electromagnetic field induction models \cite{Sharma2005} \cite{Garrido2008} and particle density models \cite{Blanken2018} \cite{Boyer2013} \cite{Graber2021}. Particle density models are widely used to model the key phenomena and complex processes in tokamak plasma reactions. In \cite{Blanken2018}, the author develops a control-oriented model based on the schematic 2D representation of the tokamak components shown in Figure \ref{fig3}. The details of tokamak operation can be found in \cite{Blanken2018}. Although this model is control-oriented, it remains very complex for predicting the tritium inventory. This is why we propose a schematic simplification to see the tokamak as a chemical reactor with two heterogeneous phases (Figure \ref{fig3}). The plasma, where particle reactions occur, and the gas phase that covers the neutral vacuum and wall components region. Pellet injection fuels the core of the plasma, while gas puffing will fuel the edge of the plasma. Gas injection provides the initial gas feed to initiate the pulse. Notice that such simplification will help us predict isotopes' accountability, but this representation does not correspond to the actual operation of a tokamak fusion reactor, because some phenomena are not captured by this model. For instance, the dynamics of the Scrape-Off Layer (SOL). Using the process diagram simplification, we can formulate isotope balances within the plasma phase to track the isotopes' reactions as they are simulated in chemical reactors. Once the particles cool, they become gas molecules. Assuming there is no mass leakage, the flow should be equal to the exhaust gas flow. We propose isotopes' molar balances in the plasma phase to obtain particle composition dynamics. Considering the following plasma reactions \cite{Motevalli2021}:

\begin{eqnarray}
D+T \longrightarrow \ce{^4He} +n, \\
D+D \longrightarrow \ce{^3He} +n, \\
D+D \longrightarrow T+H, \\
D+\ce{^3He} \longrightarrow \ce{^4He}+H.
\end{eqnarray}

Consider the isotopes' mol balances on plasma phase:

\begin{equation}
\frac{dN_{j}}{dt}=\dot{n}_{PIj}+\dot{n}_{GPj}+\dot{n}_{GIj}-\dot{n}_{exhj} \pm r V_{tok},
\end{equation}

\noindent with the subscript $j=[T,D,H,\ce{^3He},NG]$ considering that component moles can be calculated using molar fraction ($ \dot{n}_j=x_{j} \dot{n}$), and molar composition ($ \dot{N}_j=C_{j} V$ and $ \dot{n}_j=C_{j} F$). The molar particle composition dynamics are:
\\

\noindent Tritium composition:
\begin{equation}
\frac{dC_T}{dt}=\frac{\dot{n}_{PI}}{V_{tok}}x_{PI,T}+\frac{\dot{n}_{GP}}{V_{tok}}x_{GP,T}+\frac{\dot{n}_{GI}}{V_{tok}}x_{GI,T}-\frac{F_{exh}}{V_{tok}}C_T-r_1+r_3.
\end{equation}
Deuterium composition:
\begin{equation}
\frac{dC_D}{dt}=\frac{\dot{n}_{PI}}{V_{tok}}x_{PI,D}+\frac{\dot{n}_{GP}}{V_{tok}}x_{GP,D}+\frac{\dot{n}_{GI}}{V_{tok}}x_{GI,D}-\frac{F_{exh}}{V_{tok}}C_D-r_1-2r_2-2r_3-r_4.
\end{equation}
Protium composition:
\begin{equation}
\frac{dC_H}{dt}=\frac{\dot{n}_{PI}}{V_{tok}}x_{PI,H}+\frac{\dot{n}_{GP}}{V_{tok}}x_{GP,H}+\frac{\dot{n}_{GI}}{V_{tok}}x_{GI,H}-\frac{F_{exh}}{V_{tok}}C_H+r_3+r_4.
\end{equation}
Helium-3 composition:
\begin{equation}
\frac{dC_{\ce{^3He}}}{dt}=\frac{\dot{n}_{PI}}{V_{tok}}x_{PI,{\ce{^3He}}}+\frac{\dot{n}_{GP}}{V_{tok}}x_{GP,{\ce{^3He}}}+\frac{\dot{n}_{GI}}{V_{tok}}x_{GI,{\ce{^3He}}}-\frac{F_{exh}}{V_{tok}}C_{\ce{^3He}}-r_4+r_2.
\end{equation}
Inert Noble Gas composition:
\begin{equation}
\frac{dC_{NG}}{dt}=\frac{\dot{n}_{PI}}{V_{tok}}x_{PI,{NG}}+\frac{\dot{n}_{GP}}{V_{tok}}x_{GP,{NG}}+\frac{\dot{n}_{GI}}{V_{tok}}x_{GI,{NG}}-\frac{F_{exh}}{V_{tok}}C_{NG}+r_1+r_4.
\end{equation}

The reactions will follow the next empirical equations:

\begin{eqnarray}
r_1=k_1C_DC_T, \\
r_2=k_2C_D^2, \\
r_3=k_3C_D^2, \\
r_4=k_4C_DC_{\ce{^3He}}.
\end{eqnarray}

Plasma reaction kinetics coefficients are calculated approximately using Maxwellian fusion reactions reactivity (\cite{Motevalli2021}; \cite{Boyer2015}; \cite{Hively1977}). Notice we are using volumetric composition as a variable, rather than molar fraction. This is because we want to propose a simple model, and reaction kinetics can be easily calculated from volumetric composition.  

No reactions take place in the gas phase. However, it is necessary to calculate the gas composition using information from the plasma-phase particle composition. Specifically, given the isotopic fractions, the molecular composition of the hydrogen gas mixture $[H_2,D_2,T_2,HD,HT,DT]$ is determined assuming that the particles reach isotopic equilibrium in the gas phase.

To calculate the isotopic equilibrium, the following hydrogen isotope reactions are considered:

\begin{eqnarray}
H_2+D_2 \longrightarrow 2HD, \\
H_2+T_2 \longrightarrow 2HT, \\
D_2+T_2 \longrightarrow 2DT.
\end{eqnarray}

Equilibrium constants are given by:

\begin{equation}
K_{HD}=\frac{C_{HD}^2}{C_{H_2}C_{D_2}}=a_{HD} \cdot exp \left( -\frac{b_{HD}}{T} \right),
\end{equation}

\begin{equation}
K_{HT}=\frac{C_{HT}^2}{C_{H_2}C_{T_2}}=a_{HT} \cdot exp \left( -\frac{b_{HT}}{T} \right),
\end{equation}

\begin{equation}
K_{DT}=\frac{C_{DT}^2}{C_{D_2}C_{T_2}}=a_{DT} \cdot exp \left( -\frac{b_{DT}}{T} \right).
\end{equation}

Equilibrium constant equations, together with particle balances, produce enough equations to evaluate the outflow gas concentration. These equations describe steady-state operation during the flat-top. For future research, we will consider the full operation, including ramp-up, ramp-down, and dwell phase.

\section{Dynamic analysis and control design}
\label{sec3}

In this section, the control design for the DIRL is developed using tools from dynamic systems analysis and control theory, grounded in linear algebra \cite{skogestad2005multivariable}.
Based on the nonlinear model developed in section \ref{sec2}, a linear state space model is approximated around an equilibrium point (steady-state at the flat-top phase) using the Time-Based Linearization tool from Simulink \cite{simulink2024} and Matlab \cite{matlab2024}:

\begin{eqnarray}
\dot{\textbf{x}}=\textbf{Ax}+\textbf{Bu}, \\
\dot{\textbf{y}}=\textbf{Cx}+\textbf{Du},
\end{eqnarray}

\noindent where, $\textbf{x} \in \mathbb{ R}^{43}$ corresponds to the state  vector, $\textbf{u}\in \mathbb{ R}^{8}$ is the input variable vector, and $\textbf{y} \in \mathbb{ R}^{8} $ is the output variable vector. This state-space representation is used to identify relationships between input and output variables and to gain insights into the performance of control systems under the current design \cite{skogestad2005multivariable}. The DIRL control system must fulfill the following objectives: 1). Pressure should be managed in the four vessels to maintain safe operation; 2). The protium composition that is fed should be diluted to optimize the energy uptake; and  3). The isotopic tritium-deuterium ratio should be within certain margins to manage the quality of the energy uptake and control of the fusion process. The isotopic ratio is defined as:

\begin{equation}
\mathrm{IR}=\frac{x_T}{x_D},
\end{equation}

\noindent where IR is the isotopic ratio, $x_T$ is the tritium fraction, and $x_D$ is the deuterium fraction. The IR should be maintained around 1 in the fuel and buffer vessels to achieve the maximum energy uptake.

Given that there are 8 output variables, we have selected 8 input variables. Flows from the Hydride beds are selected to control the pressure in the buffer and DT feed vessels. $\dot{n}_7$ and $\dot{n}_6$ should control the FV1 and FV2 respectively (See Figure \ref{fig2}). Deuterium source flows should control the isotopic ratio on BV, FV1, and FV2, and $\eta$ will be used to dilute the protium composition at the fuel inflow. The next section will evaluate controllability, stability, and variable interactions between the selected set of input and output variables. 

\subsection{Recycle effect on control design}
\label{subsec05}

We use the metal foil and vacuum pump device separation yield ($\eta$) to characterize different steady states. We consider values from 0 to 0.8 with a step size of 0.1. Table 1 presents the results for different steady-states as a function of $\eta$. The fast recirculation of hydrogen gas can minimize the required tritium in the TFC. Therefore, less tritium is necessary from the storage. Nevertheless, there is a possibility of protium buildup, and conditions may emerge, such as instability, in which the control system faces operating challenges. To evaluate the recycling effect, Table \ref{tab1} presents some information as a function of the separation yield. Two process indices are provided: the protium fraction fed to the fusion reactor and the tritium used from the storage. Moreover, four control indices are provided for each steady state: Controllability is evaluated using the controllability matrix rank, stability is evaluated with the highest eigenvalue of the A matrix, and input-output interaction is evaluated using an RGA index based on the definition in \cite{Shahmansoorian2014}. The RGA index is defined as the ratio of the off-diagonal term to the original matrix, using the Frobenius norm \cite{skogestad2005multivariable}. Notice that the main purpose of the index is to evaluate how much the RGA is different from the ideal condition where every output corresponds to one input. This ideal situation is an RGA matrix equal to the identity matrix, with the input-output pairing organized along the diagonal. RGA index is defined as:

\begin{equation}
\mathrm{RGA}_I= \frac{\left|| \Lambda-I \right|| _{F}}{\left|| \Lambda\right|| _{F}},
\end{equation}

\noindent where $\left|| \cdot \right|| _F$ is Frobenius norm. $\Lambda$ represents the RGA matrix, $I$ is the identity matrix, and $n$ represents the number of elements $(n=8)$. Notice that $n$ can be either the number of inputs or the number of outputs, since the RGA index is defined for square systems. Moreover, the best situation corresponds to an RGA index equal to zero.

\begin{table}[t]
\centering
\begin{tabular}{l c c c c c}
$\eta$ & $Rank(Co)$ & $max(\lambda)$ & RGA$_I$ & $x_{P}$ & $\dot{n}_T$ \\
0 & 25 & $8.26\times10^{-15}$ & 0.138 & $7.75\times10^{-5}$ & 0.073 \\
0.1 & 28 & $-1.21\times10^{-8}$ & 0.168 & 0.0085 & 0.066 \\
0.2 & 30 & $4.38\times10^{-9}$ & 0.208 & 0.0158 & 0.058 \\
0.3 & 30 & $1.13\times10^{-7}$ & 0.253 & 0.0223 & 0.051 \\
0.4 & 30 & $1.28\times10^{-7}$ & 0.308 & 0.028 & 0.044 \\
0.5 & 30 & $1.03\times10^{-7}$ & 0.371 & 0.0348 & 0.036 \\
0.6 & 29 & $-8.20\times10^{-8}$ & 0.400 & 0.0444 & 0.029 \\
0.7 & 29 & $-6.41\times10^{-9}$ & 0.336 & 0.0547 & 0.022 \\
0.8 & 29 & $4.50\times10^{-8}$ & 0.279 & 0.0668 & 0.015 \\
\end{tabular}
\caption{Dynamic analysis indices as a function of separation yield}\label{tab1}
\end{table}

Table \ref{tab1} shows that the rank of the controllability matrix and the largest eigenvalue do not show significant effects with respect to the separation yield. However, they show two issues for control design: 1). The controllability matrix has rank lower than 30 for 43 state-space variables, indicating that the actual inputs cannot affect all state variables. This situation poses a challenge for control design because some state variables are not affected by the input variables; identifying these variables is necessary to avoid poor control design. 2). There are eigenvalues near zero. This implies the presence of a non-decaying mode (integrator). Consequently, a marginally stable mode exists, and the proposed control should account for this mode by employing stabilizing control; otherwise, internal control instabilities may arise. Moreover, the resulting RGA index increases with increasing separation yield, indicating a stronger interaction between the input and output variables as recirculation increases. As interactions increase, a decentralized SISO control scheme becomes more challenging, and MIMO control is usually preferred to handle the interactions.

\begin{table}[t]
\centering
\begin{tabular}{l c c c c c c c c}
 & $\dot{n}_5$ & $\dot{n}_{3}$ & $\dot{n}_{4}$ & $\dot{n}_{10}$ & $\dot{n}_{11}$ & $\dot{n}_6$ & $\dot{n}_{7}$ & $\eta$ \\
$P_{FV}$ & 0 & 0 & 0 & 0 & \textcolor{red}{1} & 0 & 0 & 0 \\
$P_{BV}$ & 0 & 0 & 0.096 & \textcolor{red}{0.345} & 0 & 0.036 & 0.013 & \textcolor{red}{0.510} \\
$P_{FV1}$ & 0.001 & 0 & 0.001 & 0.005 & 0 & 0 & \textcolor{red}{0.986} & 0.007 \\
$P_{FV2}$ & 0 & -0.003 & 0.005 & 0.021 & 0 & \textcolor{red}{0.967} & 0 & 0.010 \\
IR$_{BV}$ & 0 & 0 & \textcolor{red}{0.818} & \textcolor{red}{0.182} & 0 & 0 & 0 & 0 \\
IR$_{FV1}$ & \textcolor{red}{0.999} & 0 & -0.001 & 0.001 & 0 & 0 & 0.001 & 0 \\
IR$_{FV2}$ & 0 & \textcolor{red}{1.003} & 0 & 0 & 0 & -0.003 & 0 & 0 \\
$x_P$ & 0 & 0 & 0.080 & \textcolor{red}{0.447} & 0 & 0 & 0 & \textcolor{red}{0.473} \\
\end{tabular}
\caption{Relative Gain Array - Recirculation yield $\eta=0.6$}\label{tab2}
\end{table}

Additionally, Table \ref{tab1} indicates that the RGA index is low without recirculation and decreases as the separation yield $\eta$ exceeds 0.6; however, this interaction remains very strong. It is evident that, by setting $\eta=0.6$, the maximum RGA index is obtained. Table \ref{tab2} shows the RGA that corresponds to the maximum RGA index. A relative gain element from the RGA equal to $1$ means that there is a direct effect of the input variable on the output variable. Therefore, such variables can be paired in a SISO control loop. However, the RGA shows that some input variables can affect different output variables, especially for the buffer vessel pressure ($P_{BV}$), buffer isotopic ratio (IR$_{BV}$),  and protium concentration ($x_P$), where the relative gains differ from $1$.

Table \ref{tab1} also shows that the fraction of protium that is fed to the fusion reactor increases as the separation yield ($\eta$) increases, while the tritium use from the storage decreases by a significant amount. A high concentration of protium in the plasma reduces the energy uptake from the fusion reactor. Actually, the literature recommends a protium fraction below 0.01 \cite{Day2022}. which is not the case for separation yields higher than 0.1. As a result, separation yields exceeding 0.1 raise concerns regarding protium accumulation within the GDS.

The control objectives can be difficult to achieve given the issues of controllability, stability, and complex variable interactions. Two kinds of control schemes are considered to implement: 1). A MIMO control; 2). A decentralized SISO control with appropriate input-output pairing. The next subsections will present the implementation of these methods.

\subsection{Multivariable LQR control design}
\label{subsec06}

The interaction between input and output variables can be managed by multivariable control approaches. The Linear Quadratic Regulator (LQR) controller is a linear multivariable control that can handle instability \cite{aastrom2021}. This is possible if the unstable states, like integrator states, can be controlled by the input variables. If this is the case, the process is said to be stabilizable. Nevertheless, such a condition requires that the system be controllable, which is not the case, as shown in Table \ref{tab1}. To address the issue, a reduced-order model is developed. Such an approach allows us to eliminate the state variables we cannot control. This is possible because an evaluation of the controllability matrix shows that the uncontrollable states correspond to values near zero. Such states do not require control because they remain diluted. Therefore, the uncontrollable states are not necessary to predict the complete dynamics for the output variables. By evaluating the Singular Values of the Hankel matrix, we can reduce the order from 43 to 24 states. A controllability analysis of the reduced model yields a rank of 24, guaranteeing controllability of the complete state space. Using the reduced model to design the LQR controller allows us to stabilize the process. Since we need to observe the reduced model state \textbf{x}, a Kalman filter is applied.

The LQR, with the output error treated as a state variable, is an optimal control structure that allows us to avoid off-set error \cite{aastrom2021}. Consider the space state model given by equations 37 and 38. We extend this model by considering an integrator on the output error $\dot{z}=r_y-y$, where $r_y$ is the output variable set point. The augmented model is:

\begin{equation}
\begin{bmatrix}
\dot{x}(t) \\
\dot{z}(t)
\end{bmatrix}
=
\begin{bmatrix}
A & 0 \\
- C & 0
\end{bmatrix}
\begin{bmatrix}
x(t) \\
z(t)
\end{bmatrix}
+
\begin{bmatrix}
B \\
0
\end{bmatrix}
u(t)
+
\begin{bmatrix}
0 \\
I
\end{bmatrix}
r_y(t).
\end{equation}

Then we need to calculate the following LQR cotrol law on the augmented system:

\begin{equation}
u=-K_xx-K_iz.
\end{equation}

The matrices $K_x$ and $K_i$ must be computed by solving the following optimization problem:

\begin{equation}
J=\int (\tilde{x}Q_x\tilde{x}+uR_uu)dt,
\end{equation}

\noindent where $\tilde{x}$ is the extended state vector, which includes \textbf{x} and \textbf{z}, $Q_x$ and $R_u$ are weighted matrices. This formulation finds the optimal control that minimizes the state deviation and the control effort \cite{aastrom2021}. To improve control performance, we assign weights ($q_x=10$) to the output variables related to isotopic composition and protium fraction, while assigning a lower weight ($q_x=1$) to pressure. This is considered in the $Q_x$ matrix, which is related to the state-error cost function for LQR design. Other MIMO control schemes are possible, but it is important to consider the effects of stability and controllability on the control structure design.

\subsection{Actuator positioning design}
\label{subsec07}

\begin{figure}[t]
\centering
\includegraphics[scale=0.4]{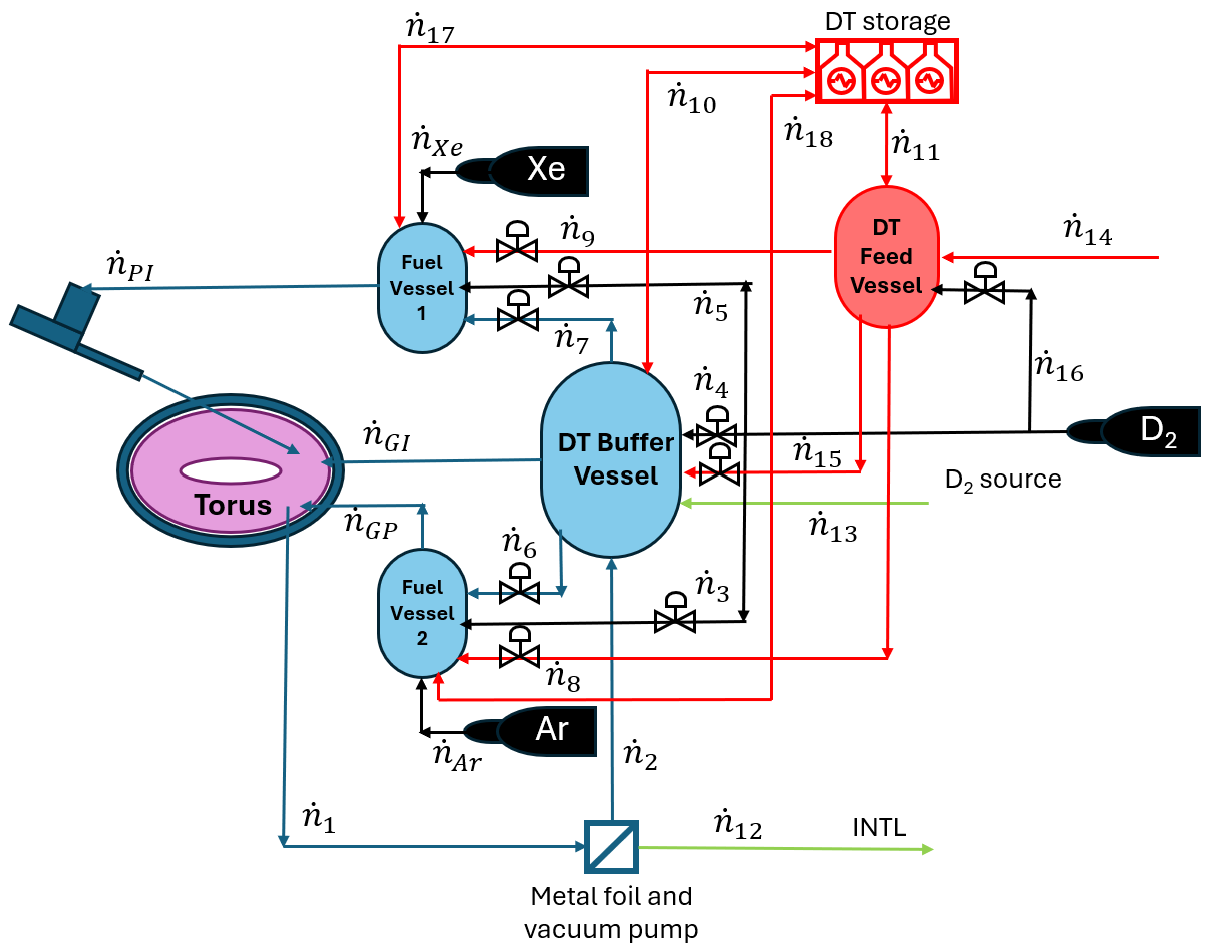}
\caption{DIRL Process Diagram with extended actuators.}\label{fig4}
\end{figure}

This section presents an analysis of actuator positioning. The set of manipulated variables selected for the previous MIMO control is extended from 8 to 14. This is possible, including one deuterium source attached to the DT feed vessel, two flows from the DT storage, and one valve that connects the DT feed vessel (See figure \ref{fig4}). Moreover, consider $\dot{n}_8$ and $\dot{n}_9$ as input variables. The purpose of including these manipulated variables is to increase input energy to manage the control challenges. We will use RGA index to design an optimization problem \cite{Shaker2013}:

\begin{equation}
\min_{B} RGA_I,
\end{equation}

subject to:

\begin{eqnarray}
b_{ij}= \{0,1\}, \sum_{i \in \Omega}b_{ij}=1, j=1,...,n_a.
\end{eqnarray}

\noindent where $\Omega$ is the subset of the state variables that can be manipulated, $n_a$ is the number of actuators, and $b_{ij}$ are the coefficients of the matrix $B$ (See section \ref{sec3}, equation 37), that are selected. Note that it is not restrictive to assume $b_{ij}=\{0,1\}$ since the system can always be normalized. Optimal placement is equivalent to determining $B$ from the integer programming problem previously depicted. We obtain a set of 8 input variables with minimal interaction for the design of SISO control systems.

\section{Results and discussion}
\label{sec4}

The dynamic analysis of the DIRL systems shows characteristics that will affect the process control. The next section presents three possible approaches to control and analyze the performance of the DIRL. The first approach is MIMO control, which can manage the interaction between input and output variables. A second approach is to design actuator positions to develop SISO control synthesis, which is favorable because it reduces variable interactions. Such an approach allows for a simpler, more economical control system. The third approach is the use of MIMO control with extended input variables. 

\subsection{Stabilizable LQR results}
\label{subsec08}

To evaluate the MIMO LQR control, we propose the following scenario. First, the control system will change the set point of the isotopic ratio of fuel vessel 1. Using a step function that goes from 1 to 1.05 at 3000 seconds. Then, we will evaluate a disturbance rejection at 6000 seconds.

\begin{figure}[t]
\centering
\includegraphics[scale=0.35]{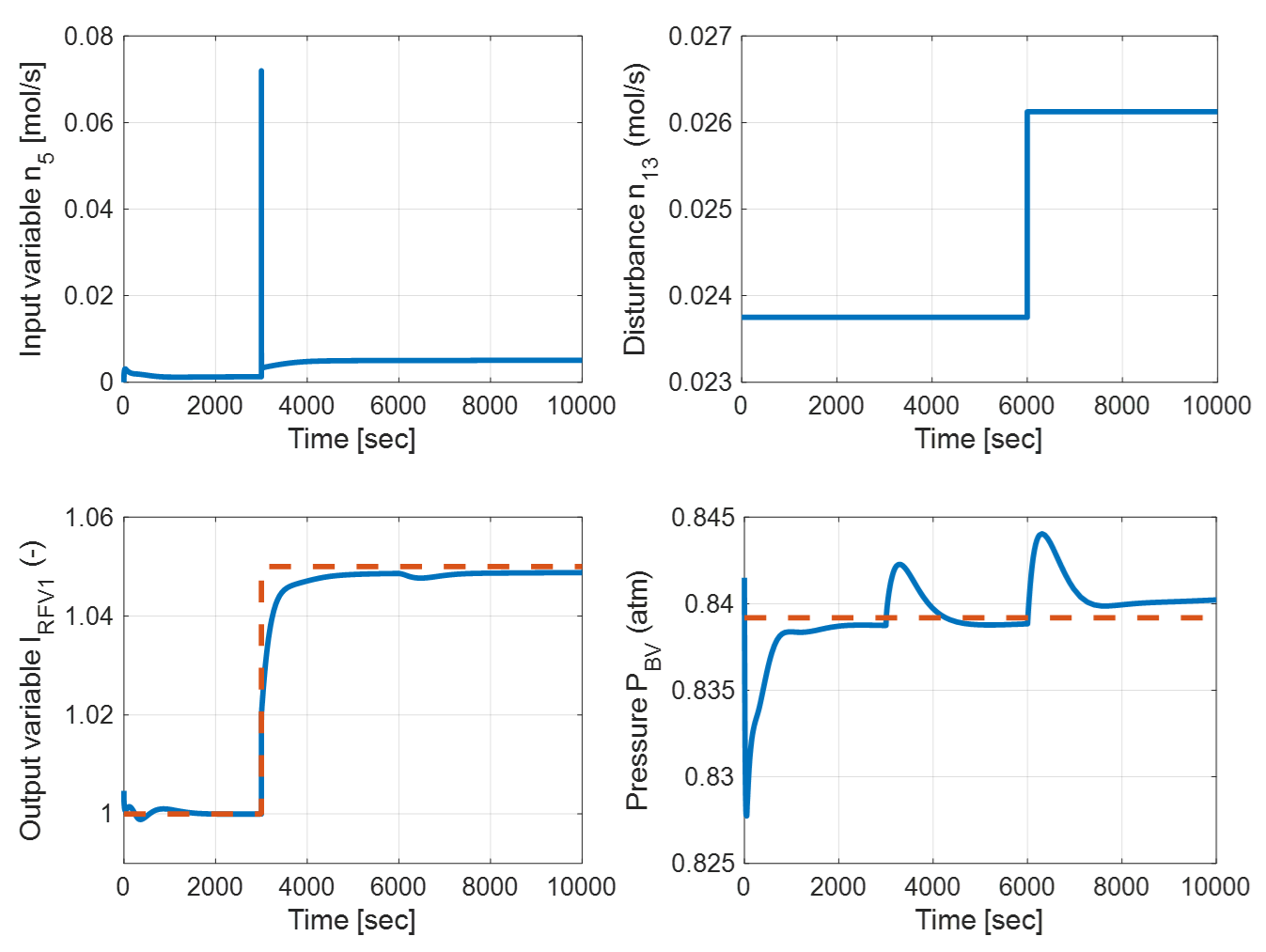}
\caption{Stabilizable LQR control performance.}\label{fig5}
\end{figure}

\begin{figure}[t]
\centering
\includegraphics[scale=0.35]{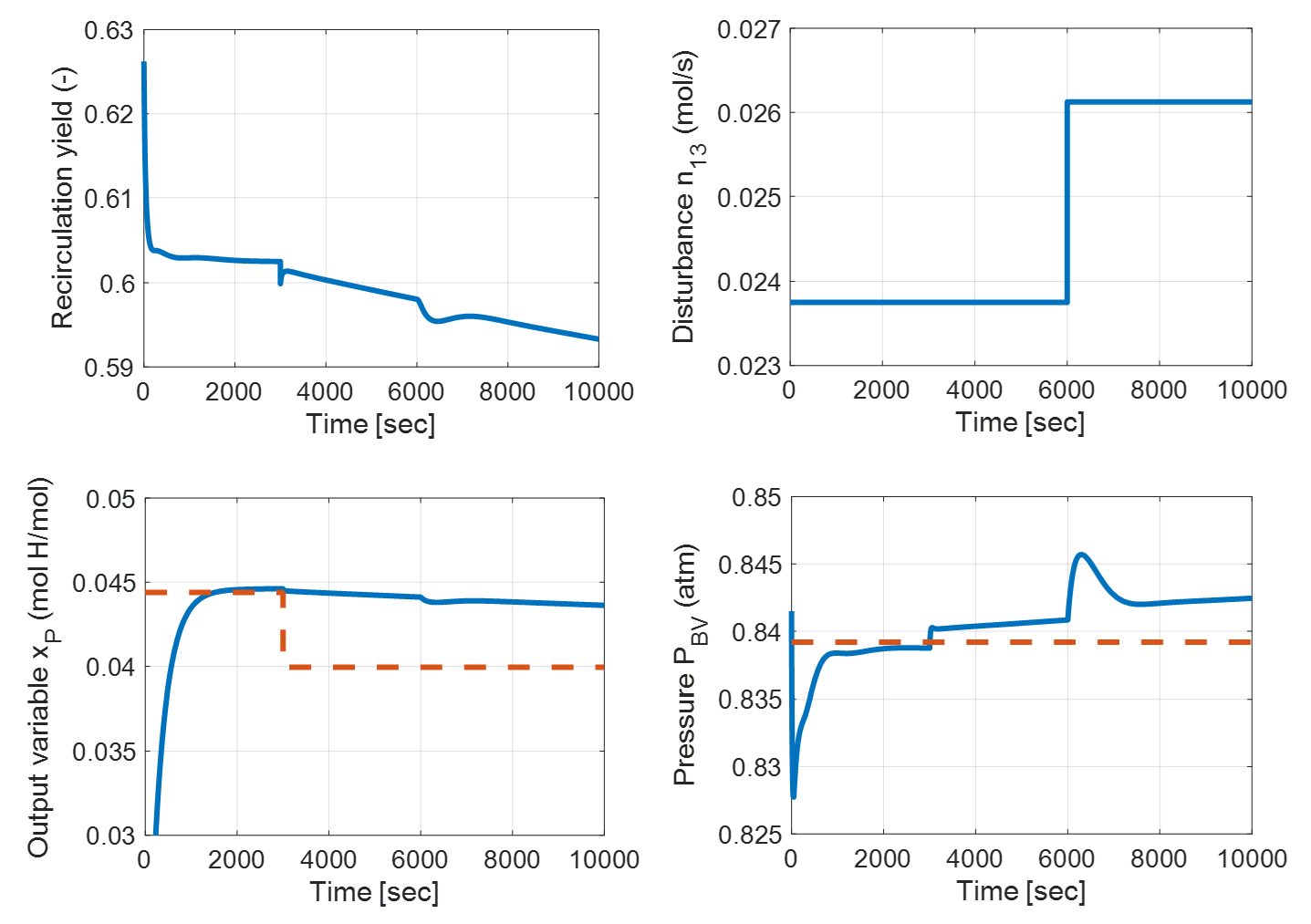}
\caption{Stabilizable LQR for protium composition dilution.}\label{fig6}
\end{figure}

Figure \ref{fig5} shows a drastic response in the input variable. This is a typical behavior in LQR control. Consequently, the input variable undergoes a sudden change to follow the reference. There is some steady-state error in both output variables. Since the Kalman filter and LQR are linear tools, the error offset may be due to nonlinear behavior or an infeasible equilibrium point. Moreover, the pressure variable controller was designed with a lower weight; as a result, it used less energy on this variable. Further evaluation, regarding the protium composition, is shown in Figure \ref{fig6}. We evaluate the controller's ability to dilute it at the gas input. This is performed using a reference step change.

In the future TFC plant, control systems need to be designed to manage the tritium inventory, gas composition, and must deal with the problem of protium buildup. Since this model considers only DIRL recirculation, further research on the inclusion of INTL and OUTL protium inventory dynamics should be conducted to analyze the impact of protium build-up and determine the operational risk. Moreover, given this challenge, is it possible to manage the gases to deliver diluted protium to the fusion reactor?

Figure \ref{fig6} shows that the LQR controller faces challenges in reducing the protium composition. This is evaluated by performing a step reducing the protium composition at 3000 seconds. In this case, it seems that the desired condition is difficult to achieve. Moreover, the pressure variable deviates from the set point. This shows that the MIMO LQR controller struggles to control the protium composition, and such additional modifications are needed to solve this issue.

\subsection{Actuator positioning design results}
\label{subsec09}

There are various methods for solving integer programming problems, such as genetic algorithms. However, since there are 3003 possible combinations of the 14 manipulated variables, complete exploration is feasible and has been performed. Optimal actuator positioning corresponds to the minimum RGA$_I$. The computed optimal result corresponds to the following flows as input variables: $\dot{n}_5, \dot{n}_{4}, \dot{n}_{11}, \dot{n}_6, \dot{n}_7, \dot{n}_8, \dot{n}_{9}, \eta$. The corresponding RGA is presented in Table \ref{tab3}.

\begin{table}[t]
\centering
\begin{tabular}{l c c c c c c c c}
 & $\dot{n}_5$ & $\dot{n}_{4}$ & $\dot{n}_{11}$ & $\dot{n}_6$ & $\dot{n}_7$ & $\dot{n}_8$ & $\dot{n}_{9}$ & $\eta$ \\
$P_{FV}$ & 0 & 0 & \textcolor{red}{0.989} & 0 & 0 & 0.010 & 0.001 & 0 \\
$P_{BV}$ & 0.008 & 0 & 0.001 & 0.016 & \textcolor{red}{0.924} & 0 & 0.054 & -0.001 \\
$P_{FV1}$ & 0.113 & 0 & 0 & 0 & 0.062 & 0 & \textcolor{red}{0.825} & 0 \\
$P_{FV2}$ & 0 & 0 & 0.012 & \textcolor{red}{1} & 0.025 & -0.016 & -0.013 & -0.009 \\
IR$_{BV}$ & 0 & \textcolor{red}{1.043} & 0 & 0 & 0 & 0 & 0 & -0.043 \\
IR$_{FV1}$ & \textcolor{red}{0.880} & 0 & 0 & 0 & 0 & 0 & 0.120 & 0 \\
IR$_{FV2}$ & 0 & 0 & -0.005 & -0.015 & 0 & \textcolor{red}{1.005} & 0.015 & 0 \\
$x_P$ & 0 & -0.043 & 0.003 & -0.001 & -0.009 & 0 & -0.002 & \textcolor{red}{1.053} \\
\end{tabular}
\caption{Relative Gain Array for optimal actuator position}\label{tab3}
\end{table}

\begin{figure}[t]
\centering
\includegraphics[scale=0.35]{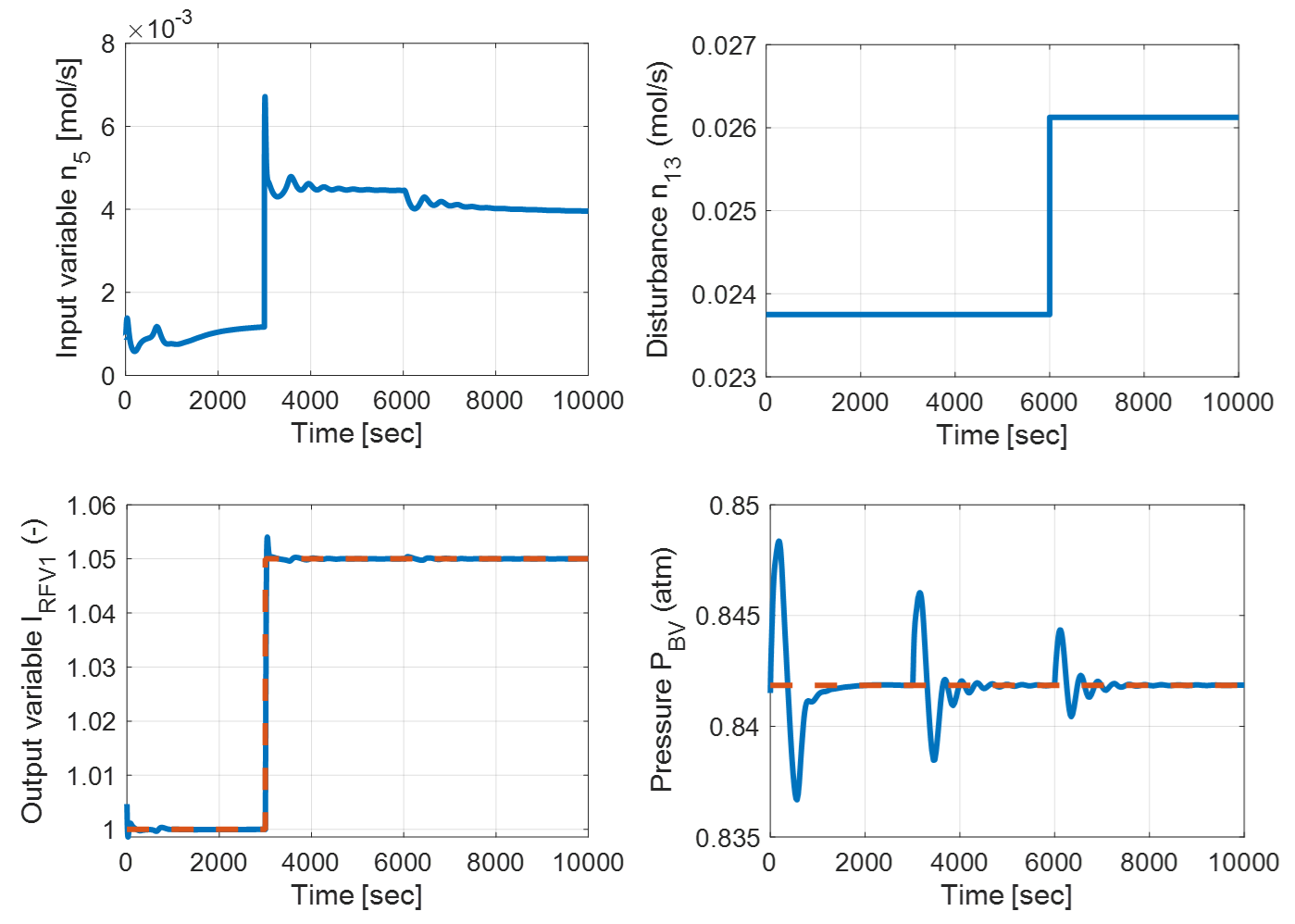}
\caption{SISO control performance.}\label{fig7}
\end{figure}

RGA results suggest that a new set of input variables is suitable for input-output pairing. This is because RGA yields values close to 1. SISO PID controllers were selected based on the RGA results, and tuning was performed using the MATLAB toolbox \cite{matlab2024}. Control performance is evaluated using a disturbance rejection scenario. Results are shown in Figure \ref{fig7}.

\begin{figure}[t]
\centering
\includegraphics[scale=0.35]{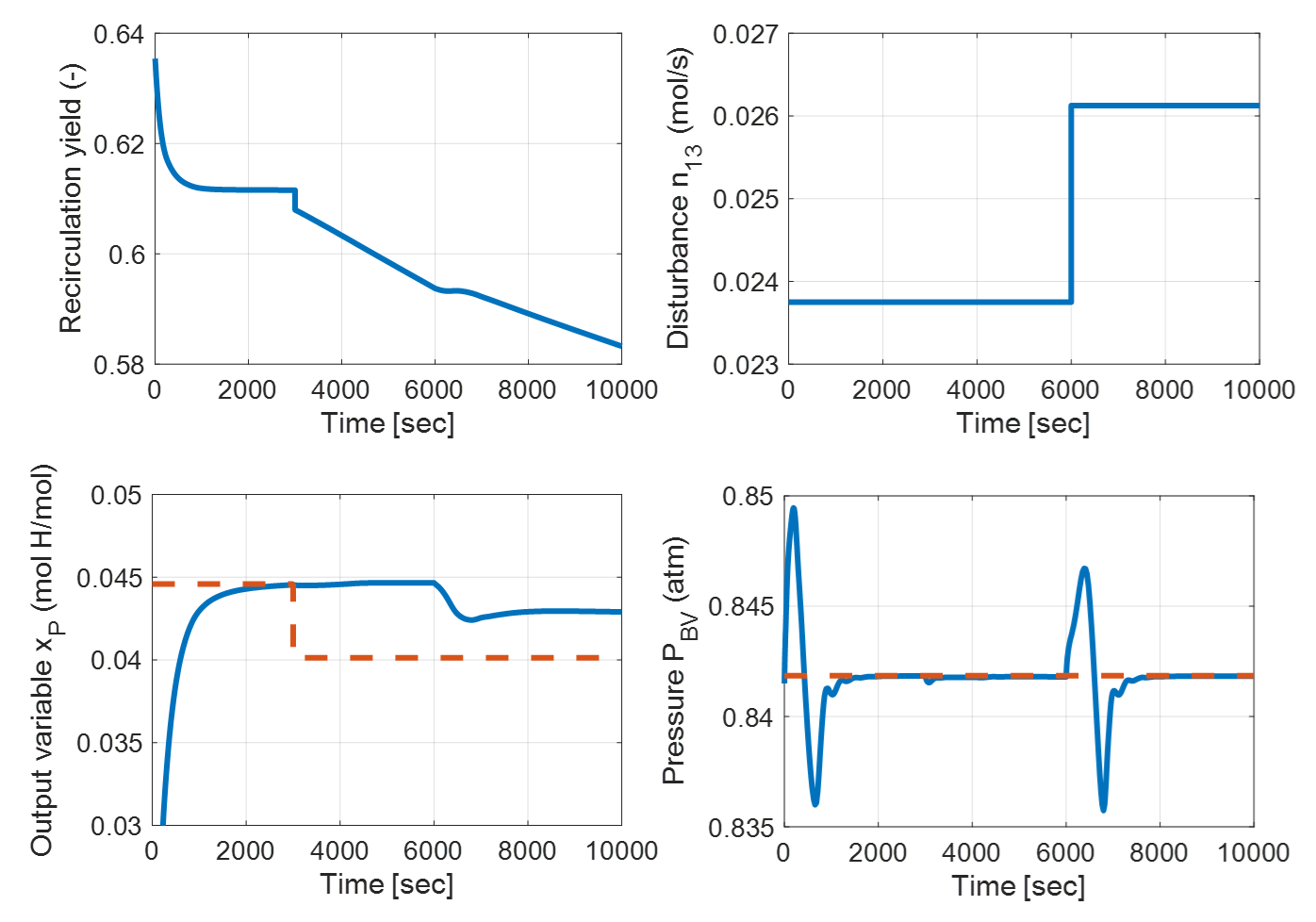}
\caption{SISO control for protium composition dilution.}\label{fig8}
\end{figure}

The SISO control performs well in the reference-tracking and disturbance scenario. PIDtune produces a highly reactive controller, leading to some oscillations in the output response. Nevertheless, the controller does not exhibit steady-state error, improving the results compared to the MIMO control case. This is because this set of input variables avoids input-output interactions. The SISO control system is evaluated with respect to the management of protium composition as the output variable. Figure \ref{fig8} shows the scenario where the protium composition set point is reduced. Although the SISO control system shows better performance, it struggles to track set points. This SISO scheme still needs further improvement to achieve the protium dilution objective described in section \ref{sec3}.

\subsection{MIMO LQR controller with extended inputs results}
\label{subsec10}

\begin{figure}[t]
\centering
\includegraphics[scale=0.35]{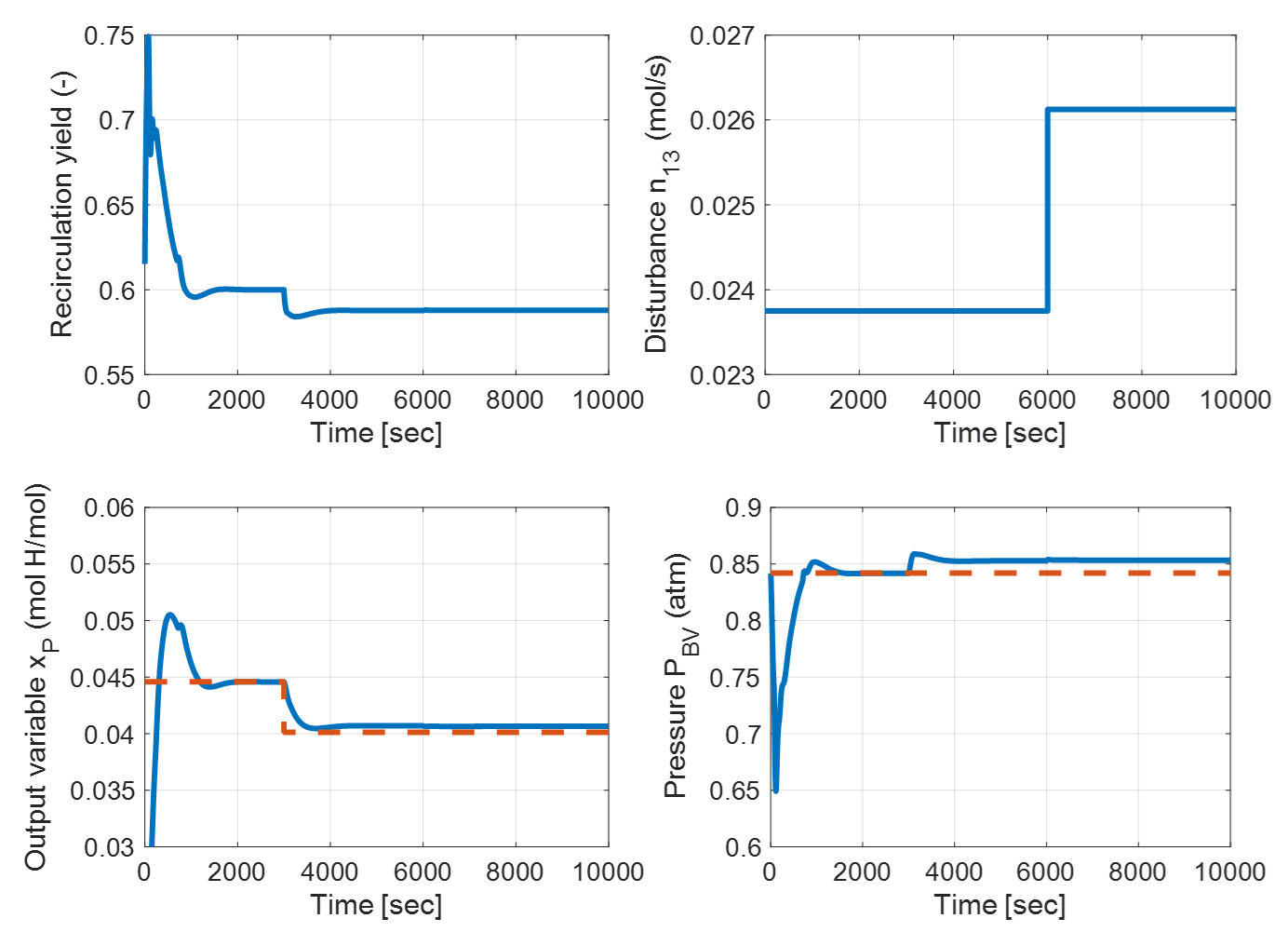}
\caption{Stabilizable LQR with more input variables applied to protium dilution.}\label{fig9}
\end{figure}

This raises a fundamental question: whether appropriate gas management strategies can effectively reduce protium composition within the system. Figure \ref{fig9} presents an LQR with extended input variables; in this case, we are using the input variables shown in figure \ref{fig3}. More input variables provide additional degrees of freedom for control, thereby improving control performance. The simulation shows that increasing the number of input variables enables control systems to manage the protium composition.

This example shows that improving the control system can enable us to achieve protium dilution, which is otherwise difficult. However, it still shows a set-point error. Further work must be done in the area of process-control co-design and advanced techniques to deal with the protium build-up problem by revisiting the DIRL design.

\section{Conclusions}
\label{sec5}

A new control-oriented dynamical model for the DIRL has been developed, focusing on the GDS system. The purpose of the model is to assess interactions among DIRL variables to propose an appropriate control system design. The dynamic model predicts pressure and composition in the vessels based on conservation balances and functional empirical equations. A dynamic evaluation was performed, identifying several issues related to controllability, stability, and input-output interactions. Moreover, protium composition in the GDS increases while the hydrogen gas recirculation through the DIRL increases. We evaluate two possible solutions: MIMO control and optimal actuator selection. Both schemes can address the issues presented. However, they cannot achieve the objectives related to diluting the protium composition. Further improvement is needed to achieve control over the protium composition. Therefore, we present MIMO LQR control with extended input variables. This scheme successfully diluted the protium composition. As a result, it is possible to achieve a lower protium composition at the inflow of the fusion reactor.

This work is the first control design analysis developed for the TFC chemical plant. This approach is evaluated under steady-state conditions during the Tokamak flat-top phase \cite{Jonas2024}, considering the fulfillment of three control objectives: 1) maintain the pressure for secure operation, 2) maintain isotopic ratios in the margins for energy quality uptake, and 3) maintain protium composition. Future work will focus on non-steady-state behavior and optimizing tritium management under stable conditions to achieve tritium self-sufficiency.

\section*{Acknowledgments}
We would like to thank Jonas Schwenzer, Yuri Igitkhanov, and Thomas Giegerich for their valuable discussions in the Eurofusion work package meetings of DEMO-TFV.
DIFFER is part of the institutes organisation of NWO.
This work has been carried out within the framework of the EUROfusion Consortium, funded by the European Union via the Euratom Research and Training Programme (Grant Agreement No 101052200 - EUROfusion). Views and opinions expressed are however those of the author(s) only and do not necessarily reflect those of the European Union or the European Commission. Neither the European Union not the European Commission can be held responsible for them.


\bibliographystyle{elsarticle-num} 
\bibliography{TFCcontrol.bib}


\end{document}